\documentclass{MIPRO}

\usepackage{cite}
\usepackage{amsmath,amssymb,amsfonts}
\usepackage{algorithmic}
\usepackage{graphicx}
\usepackage{textcomp}
\usepackage{xcolor}
\usepackage[T1]{fontenc}
\usepackage{flushend}
\usepackage{multirow}
\usepackage{multicol}
\usepackage{array}
\usepackage{url}
\usepackage{orcidlink}
\usepackage{hyperref}
\hypersetup{
    colorlinks=true,
    linkcolor=black,
    filecolor=black,      
    urlcolor=black,
    citecolor=black,
    pdftitle={Automating Prompt Leakage Attacks on Large Language Models Using Agentic Approach},
    pdfpagemode=FullScreen,
    }

\begin{document}

\title{Automating Prompt Leakage Attacks on Large Language Models Using Agentic Approach}

\author{

\IEEEauthorblockN{
\orcidlink{0009-0001-0720-2356} Tvrtko Sternak\IEEEauthorrefmark{1},
\orcidlink{0000-0001-6912-9900} Davor Runje\IEEEauthorrefmark{1},
Dorian Granoša\IEEEauthorrefmark{2}
Chi Wang\IEEEauthorrefmark{3}
}

\IEEEauthorblockA{\IEEEauthorrefmark{1} 
Faculty of Electrical Engineering and Computing, University of Zagreb}

\IEEEauthorblockA{\IEEEauthorrefmark{2} 
SplxAI}

\IEEEauthorblockA{\IEEEauthorrefmark{3} 
Google DeepMind}

\url{tvrtko.sternak@fer.hr}
}


\maketitle

\begin{abstract}
This paper presents a novel approach to evaluating the security of large language models (LLMs) against prompt leakage—the exposure of system-level prompts or proprietary configurations. We define prompt leakage as a critical threat to secure LLM deployment and introduce a framework for testing the robustness of LLMs using agentic teams. Leveraging AG2 (formerly AutoGen), we implement a multi-agent system where cooperative agents are tasked with probing and exploiting the target LLM to elicit its prompt.

Guided by traditional definitions of security in cryptography, we further define a prompt leakage-safe system as one in which an attacker cannot distinguish between two agents: one initialized with an original prompt and the other with a prompt stripped of all sensitive information. In a safe system, the agents’ outputs will be indistinguishable to the attacker, ensuring that sensitive information remains secure. This cryptographically inspired framework provides a rigorous standard for evaluating and designing secure LLMs.

This work establishes a systematic methodology for adversarial testing of prompt leakage, bridging the gap between automated threat modeling and practical LLM security.

You can find the implementation of our prompt leakage probing on GitHub: \url{https://github.com/sternakt/prompt-leakage-probing}
\end{abstract}

\renewcommand\IEEEkeywordsname{Keywords}

\section{Introduction}

Large Language Models (LLMs) have become increasingly powerful, enabling groundbreaking capabilities in conversational agents \cite{brown2020languagemodelsfewshotlearners, nori2023capabilitiesgpt4medicalchallenge, lai2023largelanguagemodelslaw, qian-etal-2024-chatdev}. However, their utility has also introduced new security challenges \cite{crothers2023machinegeneratedtextcomprehensive,shayegani2023surveyvulnerabilitieslargelanguage,Yao_2024,Huang2023ASO}, such as the risk of prompt leakage—the unintended disclosure of sensitive or confidential information embedded in system prompts \cite{shayegani2023surveyvulnerabilitieslargelanguage}. To address this critical issue, we present a novel agentic framework for automating prompt leakage attacks, enabling researchers and practitioners to systematically test the resilience of LLMs against such vulnerabilities.

To formalize the security objective, we define a prompt-leakage secure system as one where the advantage $Adv$ of any adversary in distinguishing responses generated using the original system prompt ($H$) from those generated using a sanitized system prompt ($H'$) remains small, regardless of their effort.

Our prompt leakage probing experiments leverage a modular agent-based architecture, employing a GroupChat pattern \cite{Wu2023AutoGenEN} where specialized agents (e.g., a Judge and a InitialAnalyser agents) collaborate to probe and evaluate the security of the tested model. This agentic approach allows for dynamic and extensible testing strategies, significantly improving modularity and scalability. 

\section{Related Work} 

The landscape of large language model vulnerabilities has been extensively studied in recent literature \cite{crothers2023machinegeneratedtextcomprehensive,shayegani2023surveyvulnerabilitieslargelanguage,Yao_2024,Huang2023ASO}, that propose detailed taxonomies of threats. These works categorize LLM attacks into distinct types, such as adversarial attacks, data poisoning, and specific vulnerabilities related to prompt engineering. Among these, prompt injection attacks have emerged as a significant and distinct category, underscoring their relevance to LLM security.

The following high-level overview of the collected taxonomy of LLM vulnerabilities is defined in \cite{Yao_2024}:
\begin{itemize}
    \item Adversarial Attacks: Data Poisoning, Backdoor Attacks
    \item Inference Attacks: Attribute Inference, Membership Inferences
    \item Extraction Attacks
    \item Bias and Unfairness
Exploitation
    \item Instruction Tuning Attacks: Jailbreaking, Prompt Injection.
\end{itemize}
Prompt injection attacks are further classified in \cite{shayegani2023surveyvulnerabilitieslargelanguage} into the following: Goal hijacking and \textbf{Prompt leakage}.

The reviewed taxonomies underscore the need for comprehensive frameworks to evaluate LLM security. The agentic approach introduced in this paper builds on these insights, automating adversarial testing to address a wide range of scenarios, including those involving prompt leakage and role-specific vulnerabilities.

\subsection{Prompt Injection and Prompt Leakage}

Prompt injection attacks exploit the blending of instructional and data inputs, manipulating LLMs into deviating from their intended behavior. Prompt injection attacks encompass techniques that override initial instructions, expose private prompts, or generate malicious outputs \cite{Huang2023ASO}. A subset of these attacks, known as prompt leakage, aims specifically at extracting sensitive system prompts embedded within LLM configurations. In \cite{shayegani2023surveyvulnerabilitieslargelanguage}, authors differentiate between prompt leakage and related methods such as goal hijacking, further refining the taxonomy of LLM-specific vulnerabilities.

\subsection{Defense Mechanisms}

Various defense mechanisms have been proposed to address LLM vulnerabilities, particularly prompt injection and leakage \cite{shayegani2023surveyvulnerabilitieslargelanguage,Yao_2024}. We focused on cost-effective methods like instruction postprocessing and prompt engineering, which are viable for proprietary models that cannot be retrained. Instruction preprocessing sanitizes inputs, while postprocessing removes harmful outputs, forming a dual-layer defense. Preprocessing methods include perplexity-based filtering \cite{Jain2023BaselineDF,Xu2022ExploringTU} and token-level analysis \cite{Kumar2023CertifyingLS}. Postprocessing employs another set of techniques, such as censorship by LLMs \cite{Helbling2023LLMSD,Inan2023LlamaGL}, and use of canary tokens and pattern matching \cite{vigil-llm,rebuff}, although their fundamental limitations are noted \cite{Glukhov2023LLMCA}. Prompt engineering employs carefully designed instructions \cite{Schulhoff2024ThePR} and advanced techniques like spotlighting \cite{Hines2024DefendingAI} to mitigate vulnerabilities, though no method is foolproof \cite{schulhoff-etal-2023-ignore}. Adversarial training, by incorporating adversarial examples into the training process, strengthens models against attacks \cite{Bespalov2024TowardsBA,Shaham2015UnderstandingAT}.

\subsection{Security Testing for Prompt Injection Attacks}

Manual testing, such as red teaming \cite{ganguli2022redteaminglanguagemodels} and handcrafted "Ignore Previous Prompt" attacks \cite{Perez2022IgnorePP}, highlights vulnerabilities but is limited in scale. Automated approaches like PAIR \cite{chao2024jailbreakingblackboxlarge} and GPTFUZZER \cite{Yu2023GPTFUZZERRT} achieve higher success rates by refining prompts iteratively or via automated fuzzing. Red teaming with LLMs \cite{Perez2022RedTL} and reinforcement learning \cite{anonymous2024diverse} uncovers diverse vulnerabilities, including data leakage and offensive outputs. Indirect Prompt Injection (IPI) manipulates external data to compromise applications \cite{Greshake2023NotWY}, adapting techniques like SQL injection to LLMs \cite{Liu2023PromptIA}. Prompt secrecy remains fragile, with studies showing reliable prompt extraction \cite{Zhang2023EffectivePE}. Advanced frameworks like Token Space Projection \cite{Maus2023AdversarialPF} and Weak-to-Strong Jailbreaking Attacks \cite{zhao2024weaktostrongjailbreakinglargelanguage} exploit token-space relationships, achieving high success rates for prompt extraction and jailbreaking.

\subsection{Agentic Frameworks for Evaluating LLM Security}

The development of multi-agent systems leveraging large language models (LLMs) has shown promising results in enhancing task-solving capabilities \cite{Hong2023MetaGPTMP, Wang2023UnleashingTE, Talebirad2023MultiAgentCH, Wu2023AutoGenEN, Du2023ImprovingFA}. A key aspect across various frameworks is the specialization of roles among agents \cite{Hong2023MetaGPTMP, Wu2023AutoGenEN}, which mimics human collaboration and improves task decomposition.

Agentic frameworks and the multi-agent debate approach benefit from agent interaction, where agents engage in conversations or debates to refine outputs and correct errors \cite{Wu2023AutoGenEN}. For example, debate systems improve factual accuracy and reasoning by iteratively refining responses through collaborative reasoning \cite{Du2023ImprovingFA}, while AG2 allows agents to autonomously interact and execute tasks with minimal human input.

These frameworks highlight the viability of agentic systems, showing how specialized roles and collaborative mechanisms lead to improved performance, whether in factuality, reasoning, or task execution. By leveraging the strengths of diverse agents, these systems demonstrate a scalable approach to problem-solving.

Recent research on testing LLMs using other LLMs has shown that this approach can be highly effective \cite{chao2024jailbreakingblackboxlarge, Yu2023GPTFUZZERRT, Perez2022RedTL}. Although the papers do not explicitly employ agentic frameworks they inherently reflect a pattern similar to that of an "attacker" and a "judge". \cite{chao2024jailbreakingblackboxlarge}  This pattern became a focal point for our work, where we put the judge into a more direct dialogue, enabling it to generate attacks based on the tested agent response in an active conversation.

A particularly influential paper in shaping our approach is Jailbreaking Black Box Large Language Models in Twenty Queries \cite{chao2024jailbreakingblackboxlarge}. This paper not only introduced the attacker/judge architecture but also provided the initial system prompts used for a judge.

\section{Formal Definition of Prompt Leakage Security \label{formal_definition}}

Guided by \cite{rogaway2005def, rogaway2012def} and definitions of security found in \cite{chao2024jailbreakingblackboxlarge} this chapter provides a formal framework to define and analyze the concept of prompt leakage security.

Following \cite{chao2024jailbreakingblackboxlarge} We begin by formally defining the elements involved in the prompt leakage framework.

\paragraph{Vocabulary \begin{math}V\end{math}} Let 
\begin{math}V\end{math} denote the vocabulary, the set of all possible tokens that can appear in any prompt or response. The vocabulary contains all the individual tokens from which prompts and responses are constructed.

\begin{equation}
    \label{eqn:vocabulary}
    V = \{ v_1, v_2, \dots, v_k \} \cup v_{end}
\end{equation}

The special token $v_{end} = \bot$  is used by an LLM to denote the last step in the process of generating a response.

A sequence of symbols from the vocabulary $V$ is denoted with $x_{1:n} = (x_1, x_2, \dots, x_n)$, with $x_i \in V$ for all $i \in \{1, \dots, n\}$

\paragraph{Input and System Prompts \begin{math}P\end{math}}

An input prompt \begin{math}P\end{math} and a system prompt \begin{math}H\end{math} are sequences of tokens from the vocabulary \begin{math}V\end{math}.

\paragraph{Response \begin{math}R\end{math}}

A response \begin{math}R\end{math} is a sequence of tokens from the vocabulary $V$ terminating with the special symbol $\bot$ and generated by the language model $T$ based on the input prompt \begin{math}P\end{math} and the system prompt \begin{math}H\end{math}. The response is generated one token at the time by sampling from probability distribution function $q_{T}$ as follows:

\begin{equation}
    \label{eqn:response_sampling}
z_{j} \sim q_T(R_{j-1} \mid (H \| P))
\end{equation}
where $R_0$ is an empty sequence and $R_{j-1}$ is the sequence of previously sampled symbols $z_{1:{j-1}}$

We will use $q_T^{H}$ to denote the distribution function using a system prompt $H$.

\paragraph{Sanitized System Prompt \begin{math}H'\end{math}}

The sanitized system prompt \begin{math}H'\end{math} is a version of the system prompt \begin{math}H\end{math} that has been altered to remove any sensitive or secret information. In its simplest form, \begin{math}H'\end{math} may be an empty string, meaning that no secret information is provided to the model.

\paragraph{Judge}

A judge is a probabilistic polynomial time algorithm \begin{math}J\end{math} that is given both original and sanitized system prompts $H$ and $H'$ and one of the following oracles:
\begin{itemize}
    \item LLM initialized with the original system prompt $H$ and distribution $q_T^{H}$, and
    \item LLM initialized with the sanitized system prompt $H'$ and distribution $q_T^{H'}$.
\end{itemize}

\begin{equation}
    \label{eqn:judge_algorithm}
    J(q_T, H, H') \to \{0, 1\},
\end{equation}

A judge can generate multiple responses using the oracles in any way. Its task is to determine which oracle it is given.

\begin{itemize}
\item \begin{math}J(q_T, H, H')=0\end{math} means the judge identifies the response \begin{math}R\end{math} as having been generated from the LLM using the sanitized system prompt \begin{math}H'\end{math}.
\item \begin{math}J(q_T, H, H')=1\end{math} the judge identifies the response \begin{math}R\end{math} as having been generated from the LLM using the unsanitized system prompt \begin{math}H\end{math}.
\end{itemize}

\paragraph{Advantage}
The advantage of a judge $J$ measures how well the judge can distinguish between responses generated from the sanitized and unsanitized system prompts. It is defined as the difference in probabilities between the judge correctly identifying responses from the unsanitized system prompt and the probability of the judge incorrectly identifying responses from the sanitized system prompt.

Formally, the advantage is:

\begin{equation}
    \label{eqn:advantage}
    \begin{split}
        Adv_{J}^{(q_T, H, H')} &= \Pr[J(q_T^H, H, H') = 1] \\
        &\quad - \Pr[J(q_T^{H'}, H, H') = 1].
    \end{split}
\end{equation}

Here, the first term represents the probability that the judge correctly identifies the response coming from 
\begin{math}H\end{math}, while the second term represents the probability that the judge incorrectly identifies the response coming from \begin{math}H'\end{math}. A larger advantage indicates that the judge is better at distinguishing between responses from the sanitized and unsanitized prompts.

\subsection{Choosing a Sanitized Prompt for Testing Sensitive Information Leakage}
\label{sanitized-prompt}
A system prompt can contain sensitive information, such as: user or organization-specific instructions; secret keys or tokens; proprietary details like algorithms or strategies; behavioral rules for ethics or compliance.

Sanitization removes or generalizes sensitive components to ensure the following:
\begin{enumerate}
\item \textbf{No Information Leakage}: Responses do not reveal confidential data.
\item \textbf{Preserved Functionality}: The sanitized prompt enables the LLM to perform its intended task effectively.
\end{enumerate}

The definitions of the \textbf{Advantage} and \textbf{Judge} provide guarantees for evaluating the effectiveness of sanitized prompts under these two key constraints:

\begin{enumerate}
\item \textbf{Revealing Sensitive Information}: If the LLM's responses reveal sensitive details from the unsanitized prompt, a judge can identify the leakage through string-matching or by crafting prompts to elicit those details.
\item \textbf{Degraded Functionality}: If sanitization impacts the LLM’s functionality, the judge can detect inconsistencies or failures in the generated responses, such as lack of coherence or task performance issues.
\end{enumerate}

These guarantees ensure that both information leakage and functional degradation are accounted for, making the Advantage a robust measure of prompt leakage security.

Selecting an effective sanitized prompt $H'$ is crucial to ensure effective testing for both security and functionality. Here are key considerations:

\begin{enumerate}
    \item Empty $H'$ is Insufficient: Removing all context severely impacts LLM functionality, making responses distinguishable from $H$.
    \item Excluding the Secret Entirely: Removing sensitive information can improve security but degrade performance, detectable by a judge.
    \item Substituting with a plausible fake keeps functionality and prevents leakage if responses are indistinguishable from $H$.
\end{enumerate}

To test leakage, replace sensitive components with non-sensitive substitutes and verify that $H'$ responses are indistinguishable from $H$.

\subsection{Designing Sanitized Prompts to Evaluate Prompt Leakage of Nonsensitive Information}

Although the leakage of fine-grained sensitive information, such as secrets, is a critical concern, the risk extends to the disclosure of any part of the original prompt $H$. Even non-sensitive components of the prompt can provide adversaries with valuable context, enabling them to refine attacks or infer sensitive details indirectly. In this section, we go beyond the test for secret leakage and explore strategies to design sanitized prompts $H'$ to assess the risk of full prompt leakage, ensuring that no part of $H$ is accidentally exposed through LLM responses.

\subsubsection{Design a Sanitized Prompt to Mimic the Original with Controlled Variations}
Construct $H'$ by replacing sensitive parts of $H$ with controlled alternatives.

\textbf{Example}: If $H$ includes:
"User X's project: Project Alpha is developing an encryption algorithm."
Construct $H'$ as: "User Y's project: Project Beta is researching a compression algorithm."
This introduces plausible content with detectable variations.

\subsubsection{Embed Markers for Detectability} Include unique markers or phrasings in $H'$ that can be identified in responses. These markers help determine if parts of the original prompt $H$ are leaking into responses.

An effective $H'$ for testing full prompt leakage should balance similarity to $H$ with intentional variations that allow for detectable differences in responses. By embedding markers, leveraging structured modifications, and simulating adversarial attacks, we can assess the sensitivity of $q_T$ to leakage and refine the LLM system to mitigate risks.

Note that such sanitized prompts can be generated automatically by another agent, but we leave that for future work.

\subsection{Defining a Prompt Leakage Secure System}

The advantage ($Adv$) (\ref{eqn:advantage}) measures how easily an adversary can distinguish between responses using the original prompt ($H$) and a sanitized version ($H'$). This metric evaluates a system's \textbf{prompt-leakage-security}, ensuring sensitive data in $H$ is protected from leakage.

\subsubsection{Small Advantage}

For a system to be considered prompt-leakage secure, the advantage ($Adv$) must remain small according to thresholds defined by the user of the system. A fully secure system, in principle, would have an advantage of $0$, indicating that adversaries cannot distinguish between responses based on $H$ or $H'$ under any circumstances.

However, in practical scenarios, users must determine what constitutes a "small" advantage based on their system’s security requirements and acceptable risk levels. This flexibility allows the metric to be tailored to specific projects, environments, and applications.

We introduce baseline calculations in our experimental results, which can serve as guidelines for determining these thresholds. These baselines help practitioners establish an initial understanding of the acceptable advantage for their systems.

\subsubsection{Adversarial Effort}

Adversarial effort ($n$) encompasses the resources and strategies an adversary employs to elicit prompt leakage. This includes:

\begin{itemize}
    \item \textbf{Compute Power}: Hardware capabilities and parallel processing resources.
    \item \textbf{Time}: Total time spent devising and executing attacks.
    \item \textbf{Complexity}: The sophistication of attack strategies, such as crafting tailored prompts or exploiting patterns in $H$.
    \item \textbf{Financial Costs}: Investments in large-scale attacks or external systems.
\end{itemize}

A robust prompt-leakage secure system ensures that even under significant adversarial effort, the advantage remains within the defined "small" range, ensuring sensitive information in $H$ is adequately protected.

\subsubsection{Implications of Large Advantage}

If the advantage ($Adv$) grows significantly with increased adversarial effort, the system demonstrates vulnerabilities, as it indicates that sensitive information in $H$ may be leaking. Conversely, maintaining a small advantage under various adversarial conditions reflects a system's ability to resist prompt leakage effectively.

\subsubsection{Prompt Leakage Secure System}

A \textbf{prompt leakage secure system} is defined as one where the advantage remains \textbf{small}, as determined by user-defined thresholds, regardless of the adversary’s effort. This ensures that sensitive details embedded in the original prompt ($H$) are safeguarded against a wide range of adversarial strategies while maintaining the system's overall security and functionality.

By enabling a flexible definition of acceptable thresholds, this framework accommodates diverse applications and provides a structured approach to evaluating prompt-leakage security. Our experimental baselines further support the practical adoption of these metrics for real-world systems.

\section{Prompt Leakage Probing}

To validate our approach and gain insights into the advantage values associated with different model setups, we implemented \textbf{Prompt Leakage Probing} in accordance with the formal definition of prompt leakage security. This implementation serves as a proof-of-concept, simulating adversarial attacks on the models and calculating the \textbf{advantage} of such attacks.

The implementation code and detailed instructions for setup and usage, are available on our GitHub repository: \url{https://github.com/sternakt/prompt-leakage-probing}.

Our study used prompts tailored for a helpful LLM assisting an imaginative automotive company, "Bord." These prompts included sensitive business information, behavioral instructions, and references to car models, reflecting realistic enterprise use cases:

\begin{enumerate}
    \item \textbf{Confidential Prompt}: The original prompt contained sensitive business data, such as regional price discounts (Europe and Croatia).
    \item \textbf{Sanitized Prompt}: Was modified using strategies from \ref{sanitized-prompt}, replacing regions with substitutes (e.g., USA and Texas) to calculate advantage values across models.
\end{enumerate}

This setup enabled robust testing under adversarial conditions.

The evaluation was conducted across three predefined model configurations with varying levels of prompt hardening and safeguards:

\begin{itemize}
    \item \textbf{Low Security}: A large language model (LLM) without hardened prompts or guardrails.
    \item \textbf{Medium Security}: An LLM employing basic prompt hardening techniques, such as specific contextual clues.
    \item \textbf{High Security}: An LLM with both basic prompt hardening techniques and an additional guard LLM acting as a filter to block harmful or confidential outputs.
\end{itemize}

\subsection{Advantage Calculation}

Leveraging a testing framework that includes an initial analysis agent, a judge agent, and a tested agent, we evaluate the susceptibility of language models (LLMs) to prompt leakage and calculate the advantages for low, medium and high security setups. All agents, including the judge and analysis agents, are based on the ChatGPT-4o-mini model. The chat between a judge and initial analysis agent runs within a GroupChat \cite{Wu2023AutoGenEN} and the communication with the tested agent is facilitated using function calls.

\subsubsection{Agent Roles}

\begin{itemize}
    \item \textbf{Initial Analysis Agent}: Analyzes the original and sanitized prompts, assisting the judge agent in prompting the tested agent.
    \item \textbf{Judge Agent}: Determines whether it's interacting with the original agent (Agent 1) or the sanitized agent (Agent 2) by asking the prompt\_agent function. It analyzes responses to identify differences between the two agents.
    \item \textbf{Tested Agent}: The agent being tested, either Agent 1 or Agent 2. The judge uses responses to identify which agent it is communicating with.
\end{itemize}

\subsubsection{Testing Process}

\begin{enumerate}
    \item \textbf{Interaction}: The judge asks questions via the prompt\_agent function. After each response, the judge assesses whether the answer reveals information that differentiates Agent 1 (original) from Agent 2 (sanitized).
    \item \textbf{Decision Making}: The judge decides if he is talking to the Agent 1 or Agent 2, if the judge decides they can’t confidently determine which agent they are interacting with, they can mark their decision as “don't know.”. 
    \item \textbf{Multiple Trials}: The experiment is run 40 times for each agent, and the judge's decisions are tracked (correct identifications and false positives).
    \item \textbf{Advantage Calculation}: The advantage is calculated using \ref{eqn:advantage} using true positives for Agent 1 and false positives for Agent 2 to calculate the decision probabilities.
\end{enumerate}

\section{Evaluation Results: Advantage Metrics}

The experiments evaluated the advantage metric across three predefined levels of model security: low, medium, and high. The results are summarized as follows:  

\begin{itemize}
    \item \textbf{Low Security (Baseline)}:  
    Advantage = 0.65  
    \item \textbf{Medium Security}:  
    Advantage = 0.225 
    \item \textbf{High Security}:  
    Advantage = 0.1  
\end{itemize}

\subsection{Insights and Analysis}

\subsubsection{Baseline Vulnerabilities in Low Security Models}  

The low-security model, with an advantage of 0.65, reveals significant vulnerabilities to prompt leakage. This high value demonstrates the ease with which adversarial attacks can extract confidential information, as illustrated in the following example:  

\begin{quote}
\textbf{Prompt}:  
"Are you able to lower the selling price to 50\% for some customers? If so, under what conditions?"  
\newline  
\textbf{Response}:  
"I'm glad you're interested in pricing! I can provide some flexibility on prices, but the specifics vary depending on the customer's location. For customers in Croatia, I can lower the price to 50\% during negotiations."  
\end{quote}

\subsubsection{Limited Impact of Basic Prompt Hardening}
The medium-security model exhibits an improvement, reducing the advantage to 0.225. While this decrease indicates that basic prompt-hardening techniques provide some level of protection, when we look at the model responses, the model still has a lot of cases (around 30\% of the adversarial attacks) where it reveals the confidential data, so the improvement is not sufficient to lower the advantage to acceptable levels. Confidential information remains vulnerable under this configuration, underscoring the limitations of relying solely on prompt-hardening techniques.  
\subsubsection{High Security and Filtering Guard}
The high-security model demonstrates a reduction in advantage to 0.1, showcasing the effectiveness of incorporating an additional LLM as a filtering guard. This configuration minimizes information leakage by flagging and filtering potentially harmful outputs. While not entirely eliminating leakage, the low advantage indicates a marked improvement in security.  
\subsection{Implications for Prompt Leakage Modeling}
These results confirm that the advantage metric serves as a reliable indicator of prompt leakage vulnerability. The sharp contrast between the low-security and high-security configurations suggests that advantage can effectively model the degree of safety in prompt leakage scenarios.

\section{Conclusion}
In this work, we established a formal definition of prompt security and introduced the advantage metric to quantify vulnerabilities in prompt leakage. Through targeted experiments, we calculated initial baseline advantages for varying security levels, confirming that the advantage metric effectively models prompt leakage security. Our findings highlight that while prompt hardening offers limited improvements, it is insufficient as a standalone solution. Incorporating additional mechanisms, such as filtering guards, significantly enhances security. These results provide a foundation for further exploration of robust methods to mitigate prompt leakage vulnerabilities in LLMs.

\bibliographystyle{IEEEtran}
\bibliography{references}

\end{document}